\documentstyle[12pt]{article} 
\newlength{\widthpic}
\unitlength=\widthpic
\begin{document}

\begin{center}
{\Large{\bf AN INVESTIGATION INTO THE GEOMETRY OF SEYFERT GALAXIES
 }}\\[0.3in]C. J. Clarke$^1$, A. L. Kinney$^{1,2,3}$ \& J. E. Pringle$^{1,2}$\\
$^1$ Institute of Astronomy, University of Cambridge, Mandingley Road,
Cambridge, CB3 0HA, United Kingdom\\$^2$Space Telescope Science Institute, 
3700 San Martin Dr., Baltimore, MD 21218\\$^3$Physics and Astronomy Department, Johns Hopkins University, Homewood Campus, Baltimore, MD 21218\

\end{center}

\bigskip
\begin{large}
\noindent{\bf Abstract}
\end{large}
\smallskip

We present a new method for the statistical investigation into the
distributions of the angle $\beta$ between the radio axis and the
normal to the galactic disk for a sample of Seyfert galaxies. We
discuss how further observations of the sample galaxies can strengthen
the conclusions.  Our data are consistent with the hypothesis that AGN
jets are oriented randomly in space, independent of the position of
the plane of the galaxy. By making the simple assumption that the
Standard Model of AGN holds, with a universal opening angle of the
thick torus of $\phi_{\rm c}$, we demonstrate a statistical method to
obtain an estimate of $\phi_{\rm c}$. Our data are not consistent with
the simple-minded idea that Seyfert 1s and Seyfert 2s are
differentiated solely by whether or not our line of sight lies within
some fixed angle of the jet axis.  Our result is significant on 
the 2$\sigma$ level and can thus be considered only suggestive,
not conclusive.  A complete sample of Seyfert galaxies selected on
an isotropic property is required to obtain a conclusive result.

\bigskip
\begin{large}
\noindent{\bf 1. Introduction}
\end{large}
\smallskip

We concern ourselves here with the angle $\beta$ at which the central
accretion disk of an active galactic nucleus (as evidenced by the
direction of the central radio jet) lies relative to the host-galaxy
disk, and with the angle $\phi$ which is the angle between the
line of sight and the radio axis (see Figure 1).

In the standard picture of active galactic nuclei (Antonucci 1993), it
is the angle $\phi$ which determines whether a galaxy appears
to us as a Seyfert 1 or as a Seyfert 2. The central nucleus is
surrounded by a physically thick circular torus of molecular gas which
ensures that the very central regions of the accretion flow (the broad
line region) are visible only when the line of sight and the central
disk axis (assumed to align with the radio structure) lie within about
$\phi_{\rm c} \sim 30^o$ of each other (Osterbrock and Shaw 1988). It
is a central prediction of the model, in its simplest form, that
values of $\phi$ for Seyfert 1s should be on average smaller than
values of $\phi$ for Seyfert 2s. The estimate of $\phi_{\rm c} \sim
30^o$ given by Osterbrock and Shaw comes from an estimate of the
relative numbers of Seyfert 1 and Seyfert 2 galaxies per unit volume
of space.

  The angle $\beta$ serves as a link with the innermost workings of the
active galaxy -- either as the memory of the activity-provoking event,
or as a diagnostic for the structure of the accretion disk.  We
develop a method for evaluating the most likely distribution of such
angles based on an observed set of position angles as seen on the sky
plane, and apply our method to a data set culled from the literature.
This study is a precursor to a larger study based on complete,
unbiased samples of spiral active galaxies selected on nearly
isotropic properties.

The minimum components for activity in the nuclei of galaxies are a
central black hole plus a fueling source which feeds material onto the
black hole (BH).  The processes for bringing material close to the
core (within the innermost 10pc) of a galaxy, either for the initial
formation of the BH or to provide fuel to the BH, are of two sorts --
those that trigger the gas internal to the galaxy to collapse to the
core, and those that trigger gas external to the galaxy to collapse to
the core. Internal triggers include cooling flows, bars in galaxies,
which funnel material to the core (Shlosman et al. 1989, Thronson et
al. 1989), or fly-bys of companion galaxies (Byrd et al. 1986) which
shock the ISM of the host galaxy into collapsing (perhaps aided by the
formation of a bar).  External triggers are mergers between galaxies
which bring material to the core from outside the galaxy (Hernquist
1989, Disney et al. 1995).

Spiral galaxies have ample cold gas available to be shocked into
collapse to the core by fly-bys or bars, so that internal triggers are
viable possibilities for causing activity in spirals.  But spirals
have so much gas in the disk that the gas can mask or interfere with
the evidence for a recent merger. Merger models show that the gas
merging with the spiral galaxy at random angles dissipates the angular
momentum that is not aligned with the galaxy angular momentum, and
settles into the plane of the disk so as to be indistinguishable from
the native gas (see Walker, Mihos, \& Hernquist 1996, and Mihos \&
Hernquist 1995 and references therein).

AGN in spiral galaxies appear to show precious little correlation
between the direction of inner jets/ionization cones and the spiral
disk axis (Ulvestad \& Wilson 1984; Brindle et al. 1990; Baum,
O'Dea, de Bruyn, \& Pedlar 1993; Schmitt et al. 1997). They also show
a tendency for the broad line Seyfert 1s to point out of the galaxy
disk while the narrow line Seyfert 2s point closer to the plane of the
disk (Schmitt \& Kinney 1996). This finding may, however, not be due
to the intrinsic distribution of relative position angle, but rather
due to the manner in which the active galaxies are detected for
inclusion into samples.  Of particular interest is the zone of
avoidance discovered by Schmitt, Kinney, Storchi-Bergmann, \&
Antonucci 1997 who concluded that there is apparently a zone within a
cone of half angle $20^o$ of the spiral disk axis within which no
radio axis falls.  The active nucleus apparently knows where the
galaxy plane is and avoids close alignment with it.  Given that
fueling by both internal gas and by external gas favors co-alignment
between AGN nucleus and galaxy disk, the observed mis-alignment is
intriguing and requires a theoretical explanation. it might indicate,
for example that recently suggested ideas about radiation induced
warping of accretion disks (Pringle 1996, 1997; Maloney, Begelman \&
Pringle 1996) come into play both to randomize the directionality and
to produce the ionization cones.

In this paper we undertake a more direct statistical analysis than has
been attempted hitherto in order to try to determine the distributions
of the angle $\beta$ between the jet axis and the normal to the galaxy
disk.  The main aim of the paper is to demonstrate the feasibility of
such an analysis, and to elucidate what further observations might be
required (apart from simply acquiring a larger and more uniform
sample) in order to strengthen the conclusions. In Section 2 we
describe the data set that we use. In section 3 we set up a
description of the geometry of the situation and present our
statistical method for obtaining the $\beta$ distribution together
with some preliminary results. In Section 4 we present the conclusions
of the analysis and discuss both how further observations could be
used to enhance the results and the possible implications for
theoretical models.

\bigskip
\begin{large}
\noindent{\bf 2. The Data Set}
\end{large}

We base our analysis on the data set collected by Schmitt et al.
(1997, henceforth SKSA).  The full list of objects is presented in
Table 1 of that paper.  The sample includes all Seyfert galaxies in
the literature which have been resolved in high resolution radio maps
and which show linear radio structure as defined by Ulvestad
and Wilson (1984).  The position angles of the extended radio
structure were taken from the literature, while the position angles of
the host galaxy major axis were all verified using the Digitized Sky
Survey Plates and assuming that the galaxy disk is intrinsically
circular.  No galaxies in the sample lie within 10 degrees of edge
on. Such objects are apparently difficult to detect as Seyfert galaxies,
perhaps because in such a case the active component would be entirely
absorbed by the host galaxy disk.  Our method requires a measurement
of both $i$, the galaxy inclination, and the radio position angle.
Since 4 of the objects in the Schmitt et al. sample of 46 have no
measurement for $i$, we include 42 objects in our analysis. Of these 14
are Seyfert 1s and 28 are Seyfert 2s.

When using a sample of objects assembled from the literature, one must
always consider selection effects.  SKSA show that the Seyfert 1s and
Seyfert 2s in the sample have comparable 6cm radio fluxes.  Likewise,
the Seyfert 1s and the Seyfert 2s are hosted by galaxies of similar
morphological type.  However, Ulvestad and Wilson (1984) note that
overall, Seyfert 1 galaxies which are resolved in the radio
display shorter elongated radio structures than radio resolved Seyfert
2 galaxies.  The standard model of AGN (Antonucci 1993) would
interpret this as being due to the Seyfert 1 galaxies being those
which observed within about 30 degrees of the torus axis in order that the
broad line region can be observed. The face-on orientation would
therefore tend to foreshorten the radio emission on the sky.  But
meanwhile, such foreshortening may cause Seyfert 1 galaxies to fall
out of the sample, since the radio emission must be linear to be
included in the sample.  This is a sort of blind spot in the sample,
but it should be a randomly distributed blind spot since the Seyfert
1s that are excluded can still have their radio jet extending from any
angle out of the host galaxy (except for in the plane of the host
galaxy - but those objects do not seem to make it into the sample in great
numbers).  The exclusion of Seyfert 1s with exactly face-on tori from the 
sample should affect absolute numbers of Seyfert 1's versus Seyfert 2s, but
it should not affect the distributions of position angle overall. Thus
this might be expected to affect the $\phi$ distributions for Seyfert 1s,
but should have little impact on the
$\beta$ distribution.

Radio flux and size generally correlate with far infrared power
(Sanders et al. 1989).  This sample contains the objects among the
Seyfert galaxies which are detected and resolved in the radio. Thus
these are also the brightest and most extended in the radio.  The main
selection effect is therefore the relatively benign one of selecting
the top end of the luminosity function.  The phenomenology of the most
luminous of the Seyfert galaxies may differ from the phenomonlogy of
the less luminous. An investigation of such luminosity effects could
well be of interest, but will have to wait for deeper and higher
resolution radio data.

We note that, although we are using the same sample as SKSA.  our
approach to the data is different. SKSA, in common with previous
authors (for example, Ulvestad \& Wilson 1984, Baum et al. 1993) made
use of just the distribution of position angle $\delta$ between the
radio axis and the apparent galaxy major axis. Here we make use of the
fact that for almost all of the sample we know the galaxy inclination
(assuming that the galaxy disk is intrinsically circular) as well as
$\delta$. This enables us to undertake a more direct approach by
constraining the geometry of individual objects, as we describe
below. To take this point further we show in Figure 2 the
distributions of pairs $ (\delta, i)$ for both the Seyfert 1s and the
Seyfert 2s. Evident from the Figure the selection effect on
inclination $i$, in that near face-on and near edge-on galaxies have
been excluded form the sample. Using a KS test, the distributions in
$i$ are not found to differ between the two samples, and indeed do not
differ significantly from a uniform distribution over the sphere
within the apparent sample limits. However the $\delta$ distributions
of the Seyfert 1s and the Seyfert 2s do differ, with the probability
that the two samples are drawn from the same distribution being less
that 1 per cent. This difference can be seen clearly in the Figure,
with the Seyfert 2s predominating in the range $0^o < \delta < 30^0$.

\bigskip
\begin{large}
\noindent{\bf 3. Statistical Analysis}
\end{large}

\smallskip
\noindent{\bf 3.1 Geometry}

For each galaxy we can determine two observational parameters,
$i$ and $\delta$. The angle $i$ is the inclination of the plane of
the galaxy to the plane of the sky, or equivalently the angle
between the line of sight and the vector normal to the galaxy
plane. The angle $i$ lies in the range $0^o < i <90^o$. We use a
Cartesian coordinate system OXYZ (see Fig. 1, cf. Fig. 5 of SKSA) so that
OX lies along the apparent major axis of the galaxy disk, OY
lies along the apparent minor axis, and thus OZ is the vector
normal to the galaxy plane. In these coodinates the unit
vector in the direction of the line of sight is

$$ {\bf k}_s = ( 0, -\sin i, \cos i). \eqno (3.1.1)$$

The angle $\delta$ corresponds to the position angle between the
apparent major axis of the galaxy and the direction of the
radio jet projected onto the plane of the sky. By convention
$\delta$ is taken to lie in the range $0^o < \delta <90^o$.

For a given value of $\delta$ and $i$, the direction of the jet,
which we denote by a unit vector ${\bf k}_j$ is determined to lie on a
great circle drawn on a unit sphere centred at the origin of
our coordinate system. In the OXYZ coordinates described above
the great circle is the set of points:

$$  {\bf k}_j = ( k_{jx}, k_{jy}, k_{jz})$$
	
$$  =   ( \cos \delta  \sin \phi,$$

$$	\sin \delta  \cos i \sin \phi - \sin i \cos \phi,$$

$$	\sin \delta \sin i \sin \phi + \cos i \cos \phi), \eqno (3.1.2)$$

\noindent where $\phi$ is the angle between the vectors ${\bf k}_s$
and ${\bf k}_j$ and formally lies in the range $-180^o < \phi <
180^o$. We should also note that there is a mirror symmetry to the
problem about the apparent minor axis of the galaxy, that is about the
OYZ plane. In terms of our coordinates this translates into the
statement that reversing the direction of the OX axis leaves the
problem unchanged. Thus, formally, the sign of $k_{jx}$ is not an
observationally meaningful quantity, or in other words the jet vector
in fact lies on one of two great circles which are reflections of each
other in the OYZ plane. We have therefore simplified the discussion by
considering just one of these great circles.

Of course not every point on the great circle is physically
accessible for the reason that we assume that the observable
jet is the one that lies above the disk plane (as seen from
Earth). If we define $\beta$ as the angle the jet vector ${\bf k}_j$
makes with the disc normal OZ, then we see from Equation (3.1.2)
that

$$  \cos \beta = k_{jz}$$

$$	  =  \sin \delta \sin i \sin \phi + \cos i \cos \phi. \eqno (3.1.3)$$

\noindent Thus the only relevant values of $\phi$ are those which give
$0^o < \beta < 90^o$, or $\cos \beta > 0$. In terms of $\phi$, this
means that $\phi$ lies in the range $\phi_1 < \phi < \phi_1 +180^o$,
where $\phi_1$ is the value of

$$ \phi_1 = \tan^{-1} ( - \cot i/ \sin \delta ), \eqno (3.1.4)$$

\noindent which lies in the range $-90^o < \phi_1 <0^o$. We note that
physically, if $\phi < 0$, then the jet vector is projected
against the half of the galaxy disk which is nearest to us,
whereas $\phi > 0$ corresponds to the jet being projected against
the half of the galaxy disk which is furthest from
us. From the current data we cannot discriminate
between these two possibilities, but see, however, the discussion
in Section 5.

\smallskip
\noindent{\bf 3.2 The minimum value of $\beta$}

We have seen that a given pair of values of $i$ and $\delta$
constrains the jet vector ${\bf k}_j$ to lie on a great circle ${\bf
k}_j(\phi)$ on the unit sphere centred on the galaxy. The great circle
passes through the line of sight and makes an angle $b$ with the plane
of the galaxy (the $OXY$ plane), where

$$ b = \cos^{-1}(\cos \delta \sin i). \eqno (3.2.1)$$
This means that we can set a lower limit to the value of $\beta$,
which corresponds to the place where the great circle comes closest to the
pole OZ. This value, which we denote by $\beta_{min}$ is given by

$$ \beta_{min} = \cos^{-1} ( \sin^2 \delta \sin ^2 i + \cos^2 i)
^{1/2}. \eqno (3.2.2)$$
	
\noindent We note, however, that if we were able to distinguish between
the near and far sides of the galaxy, that is between $+\phi$
and $-\phi$, then we would be able to decide which of the two
segments of the great circle, on either side of $\phi = 0$, the
jet direction corresponds to. In that case we would have

$$ \beta_{min} = \cos^{-1} ( \sin^2 \delta \sin ^2 i + \cos^2 i) ^{1/2},$$

$$					{\rm for}\ \phi > 0$$

$$	  =	i,			{\rm for}\ \phi < 0. \eqno (3.2.3)$$
	
\noindent Furthermore, we also note that if we were able to determine if
the jet is approaching us or receding from us (that is we were able to
determine the sign of ${\bf k}_j . {\bf k}_s = \cos \phi$), then this
would further restrict the allowable range of $\phi$, (in particular
to decide whether $\phi$ is greater or less than $90^o$) and so enable
a more accurate determination of $\beta_{min}$.

However, for the purposes of this paper, in which the observational
data does not tell us which side of the galaxy disk is closer to us,
and in which we do not know whether the jet is approaching or
receding, we must take $\beta_{min}$ as given by Equation (3.2.2). In
Figure 3 we display the values of $\beta_{min}$ for the galaxies in
our sample, distinguishing between the Seyfert 1s and the Seyfert 2s.

We stress that while these results only give us one sided limits to
$\beta$, the results themselves depend only on the geometry of the
situation, and do not require any statistical modelling or additional
assumptions. In this sense the conclusions drawn from them are robust
and irrefutable. Thus we can state that, for example, more than 80
percent of the Sy2s in our sample have $\beta$ greater than $20^o$,
more than 50 percent of them have $\beta$ greater than $30^o$, and
more than 20 percent of them have $\beta$ more than $54^o$. For the
Sy1s in our sample we can state that more than 80 per cent of them
have $\beta$ greater than $17^o$, more than 50 percent of them have
$\beta$ greater than $25^o$, and more than 20 per cent have $\beta$
more than $32^o$. These findings agree with the conclusions of Schmitt
{\it et al.}(1997) who present evidence for a $20^o$ zone of avoidance
around the galactic pole.

It might be tempting to draw the conclusion from Fig. 2 that the
distributions for $\beta_{min}$ (or equivalently the distributions for
the angles $b$) differ for the Seyfert 1s and the Seyfert 2s. However,
applying a KS test to the distributions puts the probability that the
two sets of data are drawn from the same distribution as 29 percent.
This is an indication that based on the $( \delta, i)$ pairings alone
(Fig. 2) one should not expect to be able to find a significant
difference in the values of $\beta$ for Sy1s and Sy2s.
	
\smallskip
\noindent{\bf 3.3  Estimation of the P($\beta$) distribution}
						
Here we describe a method whereby, in principal, we may arrive
at an estimate for the distribution of angles $\beta$ the jet
vectors in our sample make with the galaxy disk normal. We
denote this distribution in terms of a probability
distribution P($\beta$), which is defined so that:

$$\int_{0}^{\pi/2} P(\beta) d \beta = 1. \eqno (3.3.1)$$

\noindent Thus, for example if the jets axes were randomly orientated
in space, and thus were independent of the orientation of the
galaxy disk, then we would find that P($\beta$) = sin $\beta$.

If we define the jet axis in terms of the angles $\beta$ and
$\theta$ with respect to the coordinate axes OXYZ, so that

$${\bf k}_j = (\sin \beta \cos \theta, \sin \beta \sin \theta, \cos
\beta), \eqno (3.3.2)$$

\noindent (see Fig. 1), then we note that the probability
distribution P($\beta$) corresponds to the full two-dimensional
probability distribution P($\beta$, $\theta$) integrated over the
azimuthal angle $\theta$. In order to proceed further we need to make
the assumption that the full two dimensional distribution
P($\beta$,$\theta$) is in fact independent of $\theta$. This
assumption is a simple one and is valid provided that there are no
selection effects which act preferentially to select in favour of jets
with particular values of the azimuth $\theta$. Since the angle
$\theta$ is defined in terms of coordinates which have a particular
orientation relative to our line of sight, such a selection effect
cannot be ruled out completely. Indeed in the Standard Model of AGN,
in which the difference between Seyfert 1s and Seyfert 2s depends on
whether $\phi$ is greater or less than $\phi_{\rm c}$, it is clear
that this assumption is not a good one when applied to Seyfert 1s or
Seyfert 2s separately. However, when applied to a complete sample of
Seyferts as a whole (which we do not have), the assumption may indeed
be a good one.

We begin by noting that if we were able to measure the actual value of
$\beta$ for each galaxy jet, then again assuming that there are no
significant selection effects, we could simply plot a histogram of
P($\beta$) directly from the observed sample. However we are unable to
determine an exact value of $\beta$ for each galaxy in our sample,
because we do not know where on the great circle the jet vector lies
-- that is we do not know the value of $\phi$ in Equation (3.1.2). We
can recognize, however, that if we already knew P($\beta$) (which we
are of course attempting to determine) then we would be able to
convert this information into a probability distribution for $\phi$,
that is we would be able to determine the likelihood of the jet vector
lying at various points along the great circle. This is the basis of
the method which we employ. We also note that of course, as discussed
above, if we knew in addition the sign of $\phi$ and/or the sign of
${\bf k}_j . {\bf k}_s$ = cos $\phi$, then we could further restrict
the position of the jet direction along the great circle defined by
$i$ and $\delta$ (see Section 5).

We start by assuming the we know P($\beta$), and in particular we
start from the assumption that the jets are randomly
orientated so that P($\beta$) = sin $\beta$. We call this initial
assumed distribution $P_{in}(\beta)$. This implies that along a
great circle, the assumed local surface (i.e. two-dimensional)
probability density of jet orientations is $P_{in}(\beta)/\sin\beta$. 
If the jet directions were randomly orientated, then
the probability of finding a jet vector in a particular galaxy
at a point on the unit circle between $\phi$ and $\phi + d\phi$
would be proportional to $| \sin \phi |  \ d\phi$. The necessity for the
factor $| \sin \phi |$ can be seen by noting that it enables the
method to reproduce correctly a randomly orientated jet
distribution. Thus for a given $P_{in}(\beta)$ the probability of
finding the actual jet between $\phi$ and $\phi +d\phi$ would be
$[P_{in}(\beta) / \sin \beta] | \sin \phi | \ d\phi$.

We now proceed by taking each data point and distributing it over the
unit sphere according to the above formula derived from the assumed
$P_{in}(\beta)$. Having done that we may then integrate over azimuth
on the sphere and obtain another estimate of the $\beta$ distribution,
which we call $P_{out}(\beta)$. Clearly if $P_{in}$ and $P_{out}$ are
the same (or do not differ significantly), then we have a good
estimate of the real $\beta$ distribution. If they differ
significantly, however, then we may, in principle, use the above
method iteratively\footnote{We note briefly that the simple iterative
procedure as stated here is not well posed, and that an extra
constraint such as an assumption about the smoothness of the $\beta$
distribution, or its closeness to being uniform, is likely to be
required.}.

Because we have only a small sample (even when both the Sy1s
and the Sy2s are taken together) we simplify the method by
binning the data. We bin the data in 5 bins equally spaced in
$\cos \beta$, so that if the jets are randomly orientated we
expect equal numbers of galaxies to fall into each bin. In
our estimation procedure, we note that, since we redistribute
each galaxy among the bins according to some probability
distribution, the actual number of galaxies assigned
statistically to each bin may not be a whole number. This in
turn implies that the statistical analysis of the procedure is
not straightforward. For simplicity, we adopt the use of
$\surd n$ error bars, where $n$ is the (not necessarily whole)
number in each bin, We expect this to give us a reasonable
estimate of the likely error due to sampling.

In Figure 4 we show the result obtained for $P_{out}(\beta)$, using
the input assumption that $P_{in}(\beta)$ corresponds to random jet
orientations, or in other words, the input assumption is that each jet
could lie anywhere randomly along the great circle defined by $i$ and
$\delta$. As can be seen from the Figure, there is no evidence in the
current sample that the jets orientations differ significantly from
being uniformly distributed. We note that this result does not
contradict the findings of Schmitt {\it et al.} (1997) who concluded,
in line with our findings in Section 3.2, that there is a significant
underabundance of AGN with $\beta < 20^o$. Our choice of bin size as
shown in Figure 2 means that the first bin corresponds to $0^o < \beta
< 37^o$, which results in such an effect being washed out in the
analysis.

\smallskip
\noindent{\bf 3.4  The Seyfert 1 and Seyfert 2 Subsamples}

In all of the above we have not differentiated between the AGN
according to whether they are Seyfert 1s or Seyfert 2s. However, if we
are prepared to make the hypothesis that the Standard Model of AGN is
correct, then we can make further statistical tests. We shall, for
illustration, take the strong hypothesis that AGN are either Seyfert
1s or Seyfert 2s depending on whether $\phi$ is less than or is
greater than (respectively) some canonical, and universal, value
$\phi_{\rm c}$. Given this assumption we may make an estimate for the
$P(\beta)$ distribution for the Seyfert 1s and Seyfert 2s
separately. For the Seyfert 1s we assume that each jet can be
anywhere along the great circle, with the proviso that  $\phi <
\phi_{\rm c}$. Similarly for the Seyfert 2s we make the same
assumption except that now we require $\phi > \phi_{\rm c}$. 

However, using the assumption that the value of $\phi$ is the {\it
only} difference between the Seyfert 1s and the Seyfert 2s, it follows
that the $P(\beta)$ distributions we obtain for the two subsamples
should be the same. This means that if we vary the assumed value of
$\phi_{\rm c}$ until the difference between the two $P(\beta)$ so
obtained is a minimum, then we obtain an estimate of the value of
$\phi_{\rm c}$. We have carried out this calculation, using only 3
bins in $\cos (\beta)$, since the Seyfert 1 sample only contains 14
entries. We use a simple $\chi^2$ test (with two degrees of freedom)
to see if the two distributions differ. We find that the difference
between the two distributions is a minimum at $\phi_{\rm c} \simeq
40^{\rm o}$, but that the goodness of fit is poor, with the
probability that the two distributions are different being 96
per cent (i.e approximately $2\sigma$). We show the `best
fit' in Figure 5. This implies that the simple hypothesis we have made
in order to determine an estimate for $\phi_{\rm c}$ is not
sustained. It may be that $\phi_{\rm c}$ is not a universal constant,
for example it could be a function of $\beta$, or the opening of the
torus might not be cylindrically symmetric. Alternatively, it might be
that the $\beta$ distributions for Sy1 and Sy2 galaxies are
significantly different. A possible hint in this direction comes from
the observation that for the `best fit' (and indeed the fits for all
values of $\phi_{\rm c}$) the Seyfert 1 distribution is more
concentrated towards low values of $\beta$ than is the case for
Seyfert 2s. Since this difference between the Sy1 and Sy2
distributions is not apparent in the $\beta_{min}$ distributions
(Section 3.2) it seems that it is the assumption that the Sy1s and
Sy2s are differentiated in terms of $\phi$, an assumption which is
fundamental to the Standard Model, which drives the suggestion of the
difference in the $\beta$ distributions.

Given the lack of systematic selection of the sample, the
small sample size, and the 2$\sigma$ level of significance of the
result, this result can only be called suggestive and not conclusive.

\bigskip
\begin{large}
\noindent{\bf 4. Conclusions and Discussion}
\end{large}
\smallskip

We have investigated the angle $\beta$ which the jet axis ${\bf k}_j$ makes
with the normal to the galaxy disk plane. By considering the minimum
value of $\beta$ which is allowed by the data for each galaxy, we are
able to conclude (Section 3.2) that for both Sy1s and Sy2s most of
then have angles $\beta$ which are greater than $25^o-30^o$. This is a
straightforward conclusion, independent of theoretical models, and
implies not only that jets perpendicular to the disk plane are not the
norm, but also that strict alignment is avoided.

Because the data are all derived from a projection of a
three-dimensional situation onto the two-dimensional plane of the sky,
we cannot expect to be able to reconstruct the full three-dimensional
picture without resorting to some statistical modelling. We have
demonstrated how such modelling can be carried out (Section 3.3) with
a minimum of assumptions. Our estimate of the $\beta$ distribution for
the sample as a whole (Fig. 4) shows that the distribution is
consistent with the distribution which would result from the jet being
pointed randomly with respect to the galaxy.

By making the further hypothesis that $\phi_{\rm c}$ is a universal
constant, we have been able to estimate a value for $\phi_{\rm c}$, by
choosing that value which makes the deduced $\beta$ distributions for
the Sy1s and the Sy2s most consistent with each other. We obtain the
value $\phi_{\rm c} \simeq 40^o$, but note that the two distributions
so obtained for $P(\beta)$ are significantly different.  Thus
simple-minded application of the Standard Model appears to lead to a
contradiction. The assumption that Sy1s and Sy2s are differentiated in
terms of the angle $\phi$ seems to result in general in a Sy1 $\beta$
distribution which displays significantly smaller values of $\beta$
than the Sy2 distribution.  Using a simple $\chi^2$ test to evaluate
the significance of this result, the hypothesis that the Sy1 and the
Sy2 distributions, $P(\beta)$, are the same is ruled out at the 4 per cent
level, or approximately 2$\sigma$.  Given the lack of systematic selection
criteria of the sample, and the fact that similarity is ruled out only to
2$\sigma$, this result is not conclusive but only suggestive.  A complete
sample, selected on an isotropic property, is required to obtain 
a conclusive result.

Thus, we have established the feasibility of this kind of analysis, which
provides observational measures of the $\beta$ distribution and
provides some hope that the assumptions underlying the Standard Model
for AGN might be subject to observational scrutiny. It is evident that
a larger sample would be of value. With a large enough sample, it will
also be necessary to take proper account of measurement errors in the
angles $i$ and $\delta$, and also to have a better understanding of
the selection effects present in the sample. One example of such an
effect is the probability that small values of $\phi$ are likely to be
ruled out of the sample, because severe foreshortening of the jet
means that it is impossible to measure $\delta$ with any
precision. Even without a larger sample, however, further constraints
on the geometry obtained by further observations of the sample
galaxies would strengthen the conclusions.

One such constraint would be to determine which side of the minor axis
lies physically closer to us. If we can achieve this, we can remove
the degeneracy in our knowledge of the sign of $\phi$ and in effect
improve the statistics in the sample by about a factor of two since we
can constrain the jet to lying along only a particular segment of the
great circle.  One method by which we can discover which is the nearer
edge of the galaxy is by taking long slit spectroscopy along both the
major and the minor axis.  Along the minor axis we would look for dust
effects and thus need the bluest possible waveband, where the dust has
the most effect.  By differencing the spectrum of the top minor axis
from that of the bottom minor axis we should be able to distinguish
the foreground, less-reddened side from the background, more reddened
side.  Along the major axis we would look for velocity differences,
and thus need spectral regions with distinct galactic spectral
features, which requires slightly redder spectral regions.  Then,
along with the assumption that the spiral arms are always trailing,
the velocities along extreme ends of the major axis, or the farthest
ends intercepted by a spiral arm, would easily allow us to distinguish
front from back.  The differences in velocity that we will need to
detect will be of order 600 km/s independent of the inclination $i$.
By placing the long slit on both the major and minor axes, we
determine foreground side versus background side with two independent
methods.  Both methods are more accurate for edge-on than for face-on
galaxies, but similarly, the answer is more constraining for the
edge-on than for the face-on case.

A further, though more difficult, method of constraining the geometry
would be to attempt to determine whether the radio jet is moving
towards us as opposed to away from us (not to be confused with whether
the radio jet is on the front side of the galaxy as opposed to the
back side of of the galaxy). This would then limit the sign of cos
$\phi$ and so improve our statistics by further limiting the arc on 
the great circle along which the jet can lie.  Determining whether the
jet is into the sky plane or out of the sky plane is not easy.  For
some of the stronger, more highly polarized jets, it may be possible
to look for depolarization along the length of the jet.  This method
is very effective for the grand radio jets, as demonstrated by Laing
(1988) and Garrington et al. (1988) but may be problematic for the
smaller jets found in Seyfert galaxies.  As a polarized jet is observed
further into its own local halo, the halo serves to depolarize the
jet, thus showing which end of the jet is observed through a higher
column density.

Finally we note that although further observations, both in terms of a
larger sample, and in terms of further observations of the current
sample, will provide better constraints, it is clear that we have
already demonstrated the power of this kind of approach. Even the
preliminary results which we present here, based on simple
observations of an initial sample, provide interesting implications
for theoretical interpretation of Seyfert galaxies. In particular it
may be necessary to provide a theoretical explanations (a) for why the
inner accretion disks (as evidenced by the jets) appear to have little
knowledge of the plane of the galaxy and (b) for why {\it either}
there is little difference in the $\phi$ values for Sy1s and Sy2s {\it or}
the inner disks in Sy1s are more aligned with the galaxy plane than
the disks in Sy2s. It will be of interest to see whether these results
hold up with the acquisition of further and better observational data.

\bigskip
\begin{large}
\noindent{\bf Acknowledgements}
\end{large}
\smallskip	

We would like to thank Ski Antonucci and Henrique Schmitt for discussions
and critical comments to the manuscript.  We would like to thank
Rupert Knight for his late arrival, which allowed C.J.C. to complete calculations before becoming otherwise engaged.  ALK would like to
thank the Institute of Astronomy for hospitality during a sabbatical
visit.

\bigskip
\begin{large}
\noindent{\bf References}
\end{large}
\smallskip

\noindent Antonucci, R., 1993, ARA\&A, 31, 473

\noindent Baum, S.A., O'Dea, C.P., de Bruyn, A.G., \& Pedlar, A., 1993, ApJ,
419, 553

\noindent Brindle, C., Hough, J. H., Bailey, J. A., Axon, D. J., Ward, M. J., 
Sparks, W. B. \& McLean, I. S. 1990, MNRAS, 244, 577

\noindent Byrd, G.G., Valtonen, M.J., Sundelius, B., \& Valtaoja, 1986, AA,
166, 75

\noindent Disney, M.J., Boyce, P.J., Blades, J.C., Boksenberg, A., Crane, P.
Dehargeng, J.M., Macchetto, F., Mackay, C.D., Sparks, W.B., \& Phillipps, S.,
1995, Nature, 376, 150

\noindent Garrington, S.T., Leahy, J.P., Conway, R.G., and Laing,
R.A. 1988, Nature 331, 147

\noindent Hernquist, L., 1989, Nature, 340, 687

\noindent Laing, R.A. 1988, Nature, 331, 149

\noindent Maloney, P.R., Begelman, M.C. \& Pringle, J.E., 1996, ApJ,
472, 582

\noindent Mihos, J.C., \& Hernquist, L. 1996, ApJ, 448, 41

\noindent Osterbrock, D.E., \& Shaw, R.A. 1988, ApJ, 327, 89

\noindent Pringle, J.E., 1996, MNRAS, 281, 357

\noindent Pringle, J.E., 1997, MNRAS, {\it in press}

\noindent Sanders, D.B., Phinney, E.S., Neugebauer, G., Soifer, B.T., 
and Mathews. K.  1989, ApJ 347, 29

\noindent Schmitt, H.R. \& Kinney, A.L., 1996, ApJ, 463, 498 

\noindent Schmitt, H.R., Kinney, A.L., Storchi-Bergmann, T., \&
Antonucci R., 1997, ApJ, 477, 623 (SKSA)

\noindent Shlosman, I., Frank, J., Begelman, M.C. 1989, Nature, 338, 45

\noindent Thronson, H.A., Hereld, M., Majewski, S., Greenhouse, M., 
Johnson, P., Spillar, E., Woodward, C.E., Harper,D.A., \& Rauscher,
B.J., 1989, ApJ, 343, 158

\noindent Ulvestad J.S. and Wilson A.S., 1984 ApJ 285, 439.

\noindent Walker, I.R., Mihols, J.C., \& Hernquist, L. 1995, ApJ, 460, 121

\bigskip
\begin{large}
\noindent{\bf Figure Captions}
\end{large}
\smallskip	

\noindent {\bf Figure 1.} The plane of the galaxy is the
X, Y plane, placed so that the apparent major axis
of the galaxy is the X axis. The line of sight,
$Z^{'}$, thus lies in the plane of the paper, and $i$ is the angle of 
inclination
of the galaxy.  The radio jet is represented by the arrow (and referred
to in the text as $\bf k_j$).  The angle at which the central accretion 
disk axis lies relative to the host galaxy disk is given by $\beta$.
The angle between the line of sight and the radio axis is $\phi$.
In the Standard Model, active galaxies with $\phi$ less than some canonical
value of about 30$^{\rm o}$ are observed as Seyfert 1 galaxies, while
active galaxies with $\phi$ greater than that value are observed to
be Seyfert 2 galaxies.  The position angle between the radio axis
and the apparent major axis of the galaxy is given by $\delta$.  Note
that the two observable angles are $i$ and $\delta$.  The azimuthal
angle of the radio jet is given by $\theta$.

\noindent {\bf Figure 2.} The difference in position angle on the sky (dpa)
between the jet and the apparent major axis of the galaxy (also
referred to in the text as $\delta$) is plotted against the inclination
angle $i$ of the galaxy plane (with $i = 90^o$ corresponding to an
edge-on galaxy). The Seyfert 1s are shown as filled squares and the
Seyfert 2s as crosses.

\noindent {\bf Figure 3.} The cumulative distributions of
$\beta_{min}$ are shown for both the Sy1s and the Sy2s. The
distributions are scaled so that the total number of each subsample
appear as a percentage. The Seyfert 1s are shown as filled squares and the
Seyfert 2s as crosses.

\noindent {\bf Figure 4.} The estimated $P(\beta)$ distribution
obtained for the whole sample (Sy1s and Sy2s). The bins are chosen to
have equal sizes in $1-\cos(\beta)$ so that a uniform distribution on
the sky would give equal numbers in each bin (as shown by the
horizontal line). The error bars in each bin have been simply
estimated as $\surd n$ where $n$ is the number assigned to the bin.

\noindent {\bf Figure 5.} The estimates of the $P(\beta)$
distributions for (a) the Sy1s and (b) the Sy2s which result from the
assumption that the Sy1s and Sy2s are differentiated only in terms of
whether $\phi$ is less than, or greater than, the 'best fit' value of
$\phi_{\rm c} \simeq 40^o$. As for Figure 4 the bin sizes are equal in
$1-\cos(\beta)$.

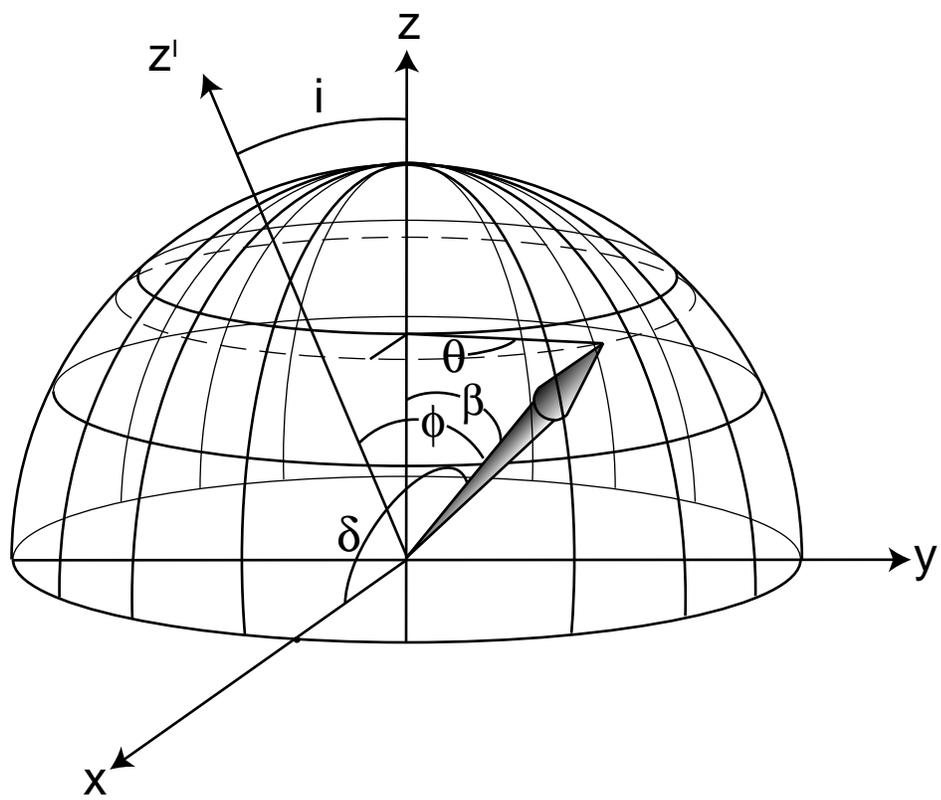
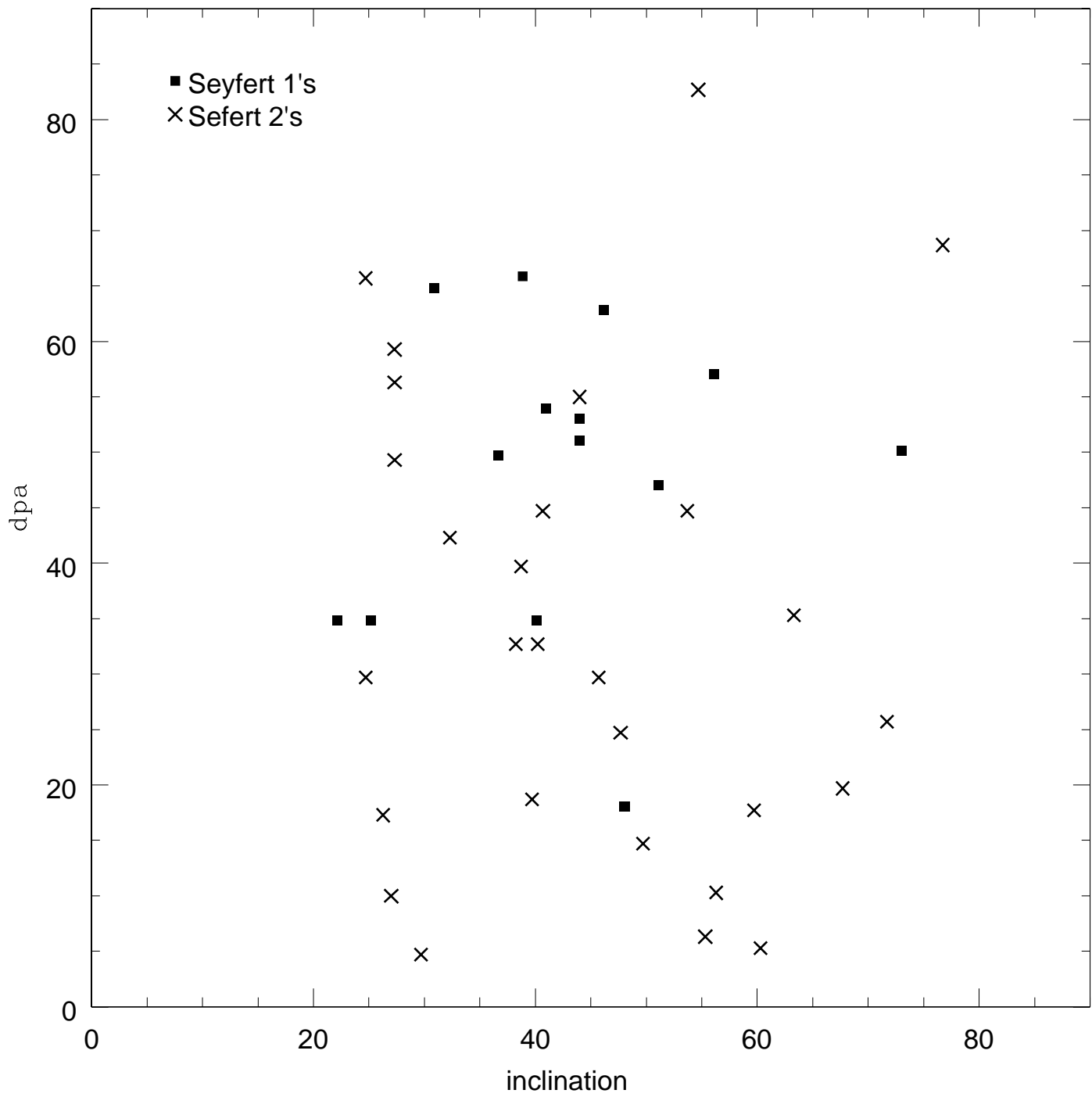
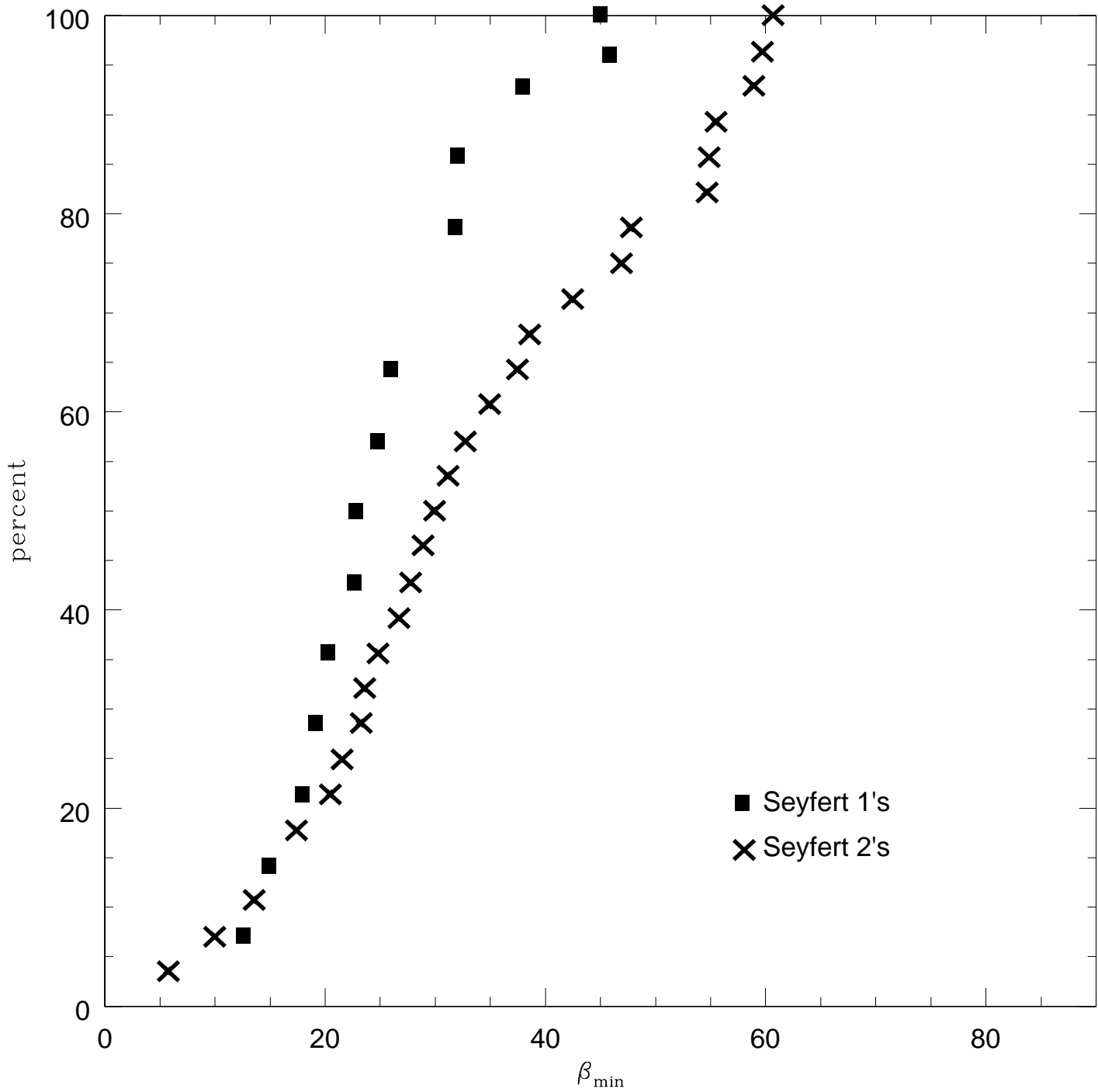
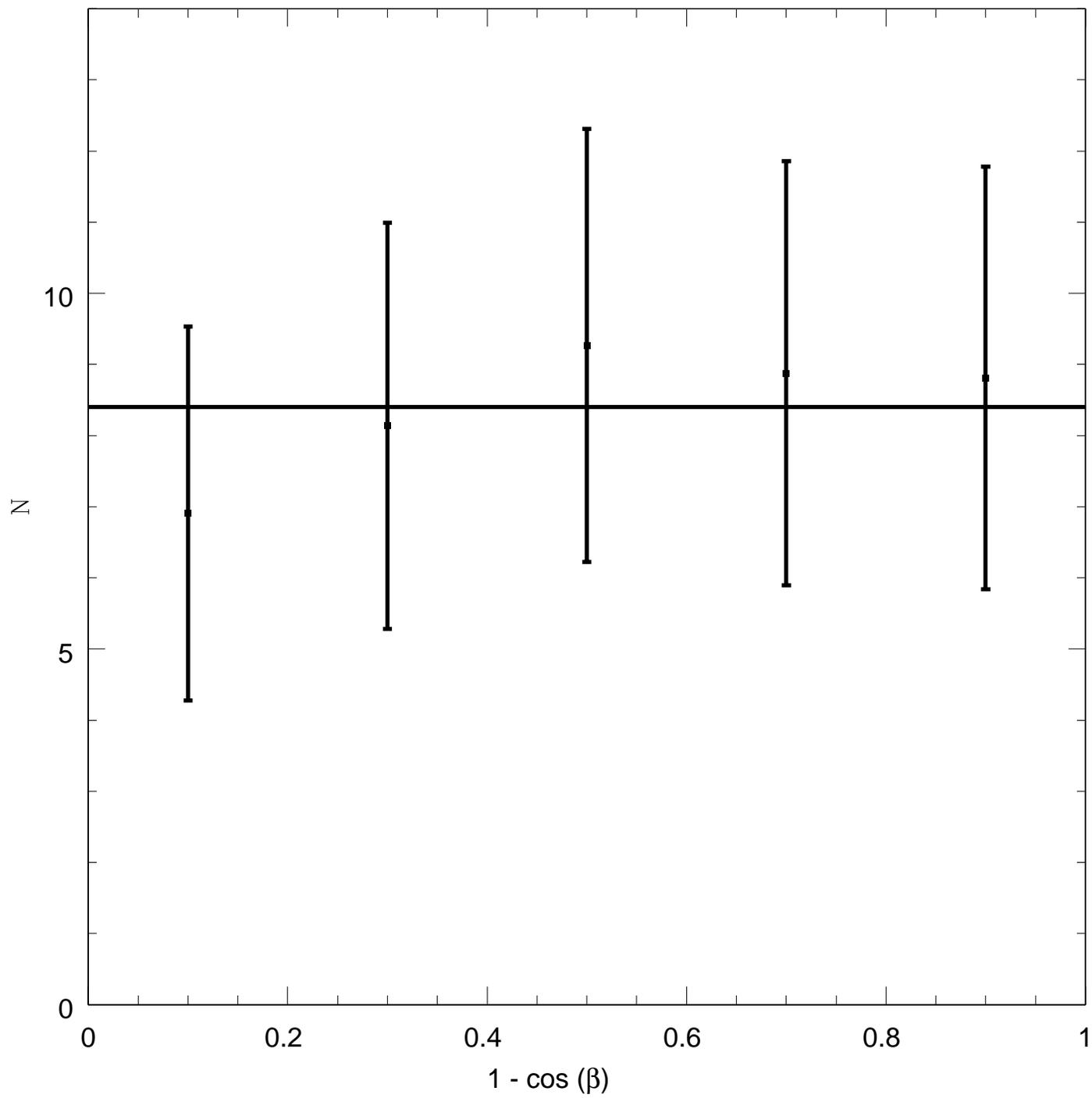
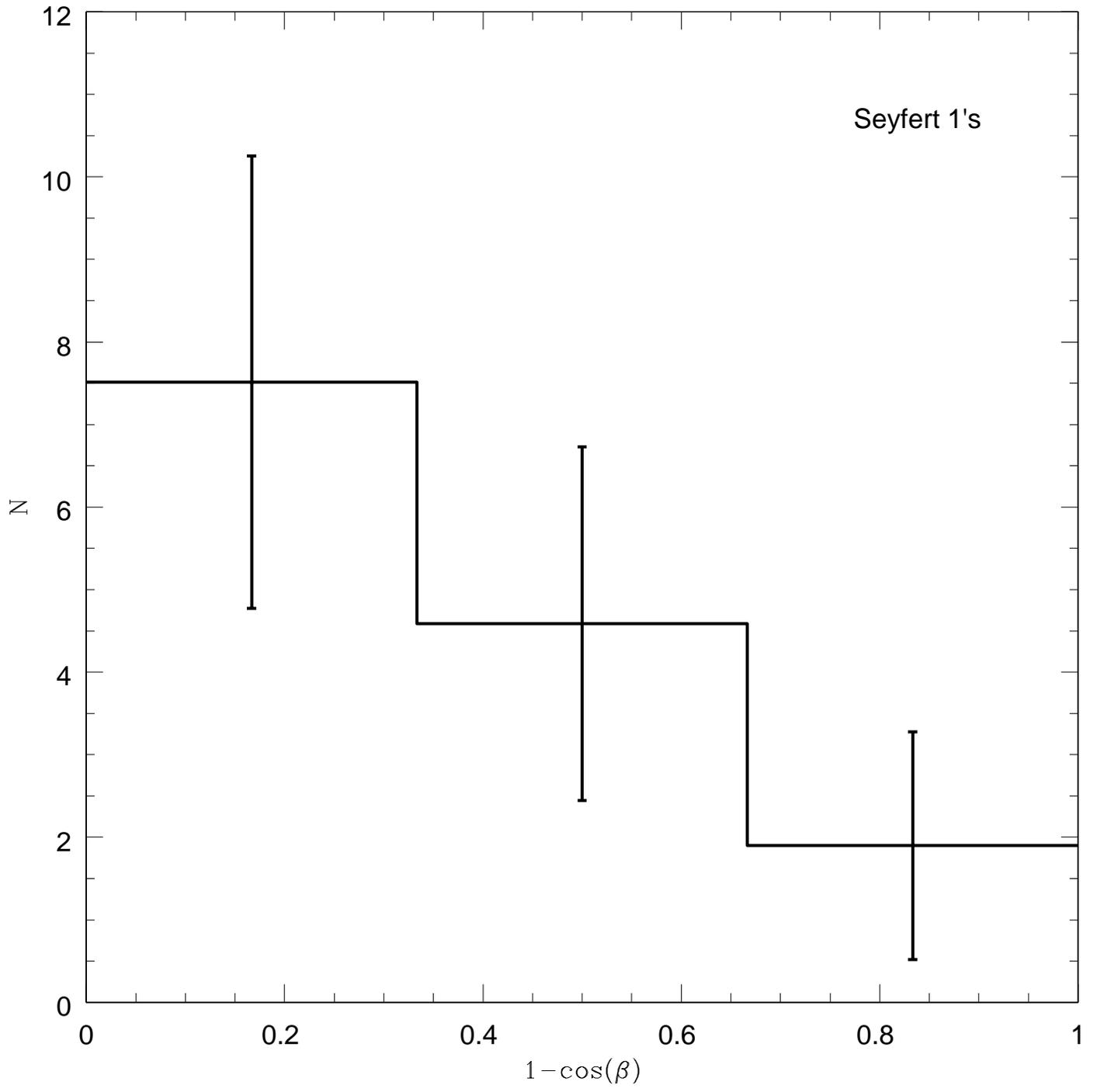
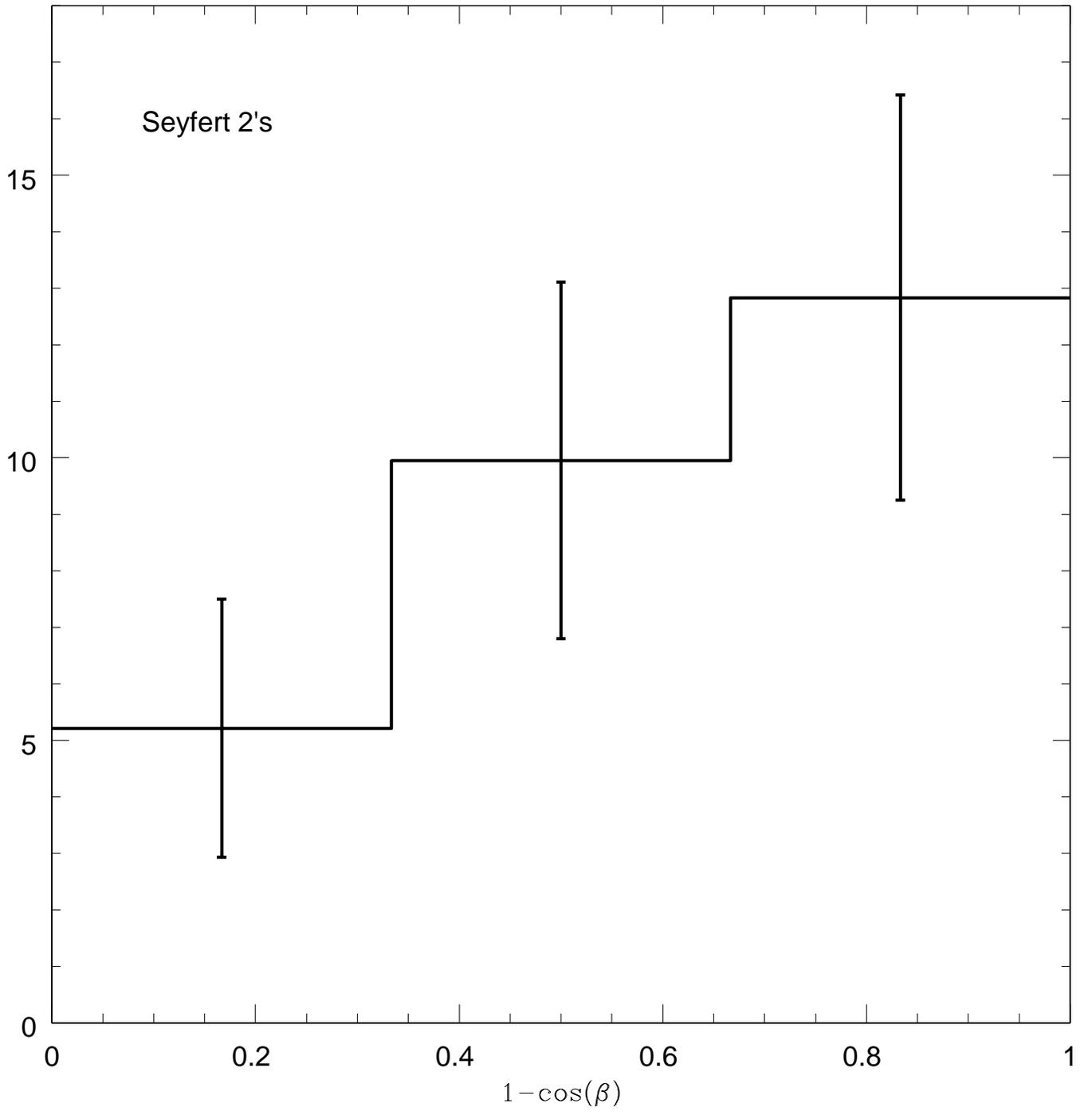
\end{document}